\documentclass[graybox]{svmult}

\usepackage{type1cm}        % activate if the above 3 fonts are
\usepackage{makeidx}         % allows index generation
\usepackage{graphicx}        % standard LaTeX graphics tool
\usepackage{multicol}        % used for the two-column index
\usepackage[bottom]{footmisc}% places footnotes at page bottom

\usepackage{newtxtext}       % 
\usepackage{newtxmath}       % selects Times Roman as basic font

\makeindex             % used for the subject index
\begin{document}

\title*{Study of $X(3872)$ and $X(3915)$ in $B\rightarrow(J/\Psi\omega)K$ at Belle}
% Use \titlerunning{Short Title} for an abbreviated version of
% your contribution title if the original one is too long
\author{Sourav Patra, Rajesh K. Maiti, Vishal  Bhardwaj}
% Use \authorrunning{Short Title} for an abbreviated version of
% your contribution title if the original one is too long
\institute{Sourav Patra \at IISER Mohali, Punjab 140306 \email{souravpatra3012@gmail.com}
\and Rajesh K. Maiti \at IISER Mohali, Punjab 140306 \email{rkumar30795@gmail.com
\and Vishal Bhardwaj \at IISER Mohali, Punjab 140306 \email{vishstar@gmail.com}}}
%
% Use the package "url.sty" to avoid
% problems with special characters
% used in your e-mail or web address
%
\maketitle

\abstract*{We present a preliminary study of $X(3872)$ and $X(3915)$ in the $B\rightarrow(J/\Psi\omega)K$ decay at Belle. This study is based on MC simulated events on the Belle detector at the KEK asymmetric-energy $e+e-$ collider.}

\abstract{We present a preliminary study of $X(3872)$ and $X(3915)$ in the $B\rightarrow(J/\Psi\omega)K$ decay at Belle. This study is based on MC simulated events on the Belle detector at the KEK asymmetric-energy $e+e-$ collider.
}

\section{Motivation}
 The $X(3872)$ was discovered by the Belle collaboration in the $B\rightarrow (J/\Psi\pi^+\pi^-)K$ decay mode. It is difficult to assign $X(3872)$ as a conventional state due to its mass near $DD^*$ threshold and narrow width (<1.2 MeV). As per the current scenario, it is expected to be an admixture of $DD^*$ molecular state and $c\bar{c}$ state. There is no  signature for the charge partner in $J/\Psi\pi^+\pi^0$ and no signature of odd charge conjugate ($C=-1$) partner in the $\eta_c\omega$ and $\eta_c\pi^+\pi^-$ decay. So, $X(3872)$ is suggested to be an iso-singlet state. In that scenario $X(3872)\rightarrow J/\psi \pi^+ \pi^-$  is isospin violating decay. Its decay to $J/\psi \omega$ is an isopsin allowed decay. It has been suggested that the ratio of $\mathcal{B}[X(3872)\rightarrow J/\psi\pi^+\pi^0\pi^-]$ to  $\mathcal{B} [X(3872)\rightarrow J/\psi\pi^+\pi^-]$ should be 30. However, BaBar collaboration has measured this ratio to be 0.8$\pm$0.3. Measuring this ratio with precision will be very useful in understanding the nature of the $X(3872)$.
%\emph{chapter.tex} 
\section{Analysis strategy}
We generated events for each of $X(3872)$ and $X(3915)$ decay using EvtGen package. Those events were simulated according to the Belle detector using GSIM. Generated MC was used to optimize and validate our study. For this analysis we reconstruct $B^\pm(B^{0})$ from $J/\Psi\omega K^\pm(K_S^0)$, where we further reconstruct $J/\psi$ from $ee$, $\mu\mu$ and $\omega$ from  $\pi^+\pi^0\pi^-$. We identify  $K_s$ in $\pi^+\pi^-$ decay and $\pi^0$ in $\gamma\gamma$ decay. Maximum unbinned likelihood fit is performed for $J/\psi\omega$ invariant mass to measure the yield for corresponding signal and backgrounds.

\subsection{Particle identification and basic selection}
 The distance of closest approach from the IP in azimuthal direction ($|dr|$) is less than   1 cm and that in horizontal direction ($|dz|$) is less than 3.5 cm. Fox-Wolfram moment ($R_2$) less than 0.5 is used to suppress continuum background events. We select the $K^\pm$ with kaon vs pion likelihood, $\mathcal{R}_K/({\mathcal{R}_K + \mathcal{R}_\pi})$ greater than 0.6 and that for $\pi^\pm$ is less than 0.4. All gamma candidates having energy more than 60 MeV and $\rm E_9/\rm E_{25}$ in ECL crystal > 0.85 are selected. $\pi^0$ candidates having mass from 123 to 147 $\textrm{MeV/c}^2$ are kept for future combination. We select $K_s^0$ having mass within [482, 524] MeV$/c^2$. We choose the mass window for selected omega from 0.7 GeV$/c^2$ to 0.85 GeV$/c^2$. $J/\psi$ candidates having mass from 3.07 to 3.13 GeV$/c^2$ for $\mu\mu$ events and from  3.05 to 3.13 GeV$/c^2$ for $ee$ events are selected. Photons within 50 mrad of each $e^\pm$ track are selected as bremsstrahlung photon to get the corrected mass and momentum for $J/\Psi$. We use two parameters: beam constrained mass ($\rm M_{\rm {bc}}$) and  $\Delta \rm E$ where $\rm M_{\rm {bc}}=\sqrt{{\rm{E}^{*2}_{cm}} - \vec{p}^{*2}_B }$ and $\Delta \rm E=\rm E^*_{beam} - \rm E^*_B$ to set the proper signal window. One should expect $\Delta E$ to peak at 0 and $\rm M_{bc}$ to peak at nominal $B$ mass. Events within $|\Delta \rm E|$ < 0.2 and $\rm M_{\rm {bc}}$ > 5.27 $\rm{GeV/c}^2$ are selected as reconstructed events for further study.
 
\subsection{Omega selection with Dalitz method}
We reconstruct $\omega$ from $\pi^+\pi^-\pi^0$. Due to its broad width and poor efficiency in $\pi^0$ reconstruction, large number of fake combinations for $\omega$ are selected. In order to avoid those fake combinations, we use Dalitz cuts. Kinematics of $\omega\rightarrow \pi^+\pi^0\pi^-$ decay can be represented in XY plane, where $\rm X=\sqrt{3}{(\rm T_{\pi^+ }-\rm T_{\pi^-})/ \rm Q}$ and  $\rm Y=(2\rm T_{\pi^0}-\rm T_{\pi^+}-\rm T_{\pi^-})/\rm Q$. Here, T is the kinetic energy of the corresponding prticle and Q implies the total kinetic energy of all three particles. We apply two concentric circular cuts centered at (0,3) in XY plane, $1.5<|\sqrt{\rm X^2 + (\rm Y-3)^2}|<3.8$, which give the maximum fake events rejection ($28.61\%$) and minimum true events rejection ($7.18\%$).

\subsection{Best candidate selection}	Multiple $B$ candidates are reconstructed  for  35$\%$ of reconstructed events. Best candidate is selected among those multiple $B$ candidates with least $\chi^2$, where
$$\chi^2 = \chi^2_V + \Big(\frac{\Delta E}{\sigma_{\Delta E}}\Big)^2 + \Big(\frac{M_{ll}-m_{J/\Psi}}{\sigma_{J/\Psi}}\Big)^2 + \Big(\frac{M_{\pi^+\pi^-\pi^0}-m_{\omega}}{\sigma_{\omega}}\Big)^2 + \Big(\frac{M_{\gamma\gamma}-m_{\pi^0}}{\sigma_{\pi^0}}\Big)^2 + \Big(\frac{M_{\pi^+\pi^-}-m_{K_S}}{\sigma_{K_S}}\Big)^2$$
Here, $\chi^2_V$ is returned $\chi^2$ from charge vertex fit and $\sigma_{\Delta E}$ is the width in $\Delta E$. $M, m, \sigma$ imply the reconstructed mass, PDG mass, mass width of the corresponding particle respectively. Truthmatched signal reconstruction efficiency using this method is $68\%$ for charged $B$ meson and $57\%$ for neutral $B$ meson.

\subsection{$\Delta \rm E $ optimization}
We optimize the $\Delta E$ window for the  candidates selected with best candidate selection to set the proper signal window for $\Delta E$. We plot figure of merit ($\rm F_{ OM}$) as a function of $\Delta E$,  where $\rm F_{OM} = {N_{sig}}/{\sqrt{N_{sig}+N_{bkg}}}$. Here, $\rm N_{sig}$ and $\rm N_{bkg}$ represent the number of signal and background events respectively. Number of  events from signal MC sample in a particular $\Delta E$ region are scaled by the branching fractions. We optimize the region $|\Delta E|$< 20 MeV as signal window.

\subsection{$\Delta E$ and $\pi^0$ mass constrain fit}
$\Delta E$ should be zero for perfectly reconstructed events. We  assume that  our $\Delta E$ resolution is not good due to problem in $\pi^0$ reconstruction. Therefore, we force $\Delta E$ to be zero by keeping $\pi^0$ invariant mass fixed. So, new $\pi^0$ momentum is shifted by a factor of $\alpha$, where $\alpha = \sqrt{(1-(1-s^2)E^2_{\pi^0}/\vec{P}_{\pi^0}^2)}$ with $s= [E_{beam}-(E_{\pi^+}+E_{\pi^-}+E_{K_s})]/E_{\pi^0}$. After performing this fit, we get $\omega$ candidate with better mass resolution.

\section{Background study}
	\begin{figure}
\centering
\begin{minipage}{.33\textwidth}
  \centering
  \includegraphics[width=1\textwidth]{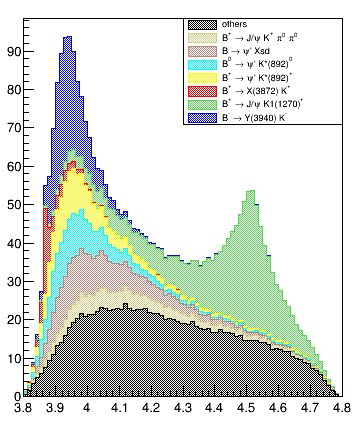}
   %\caption{$K^\pm$ events}
  %\label{fig:test1}
\end{minipage}%
\begin{minipage}{.33\textwidth}
  \centering
  \includegraphics[width=1\textwidth]{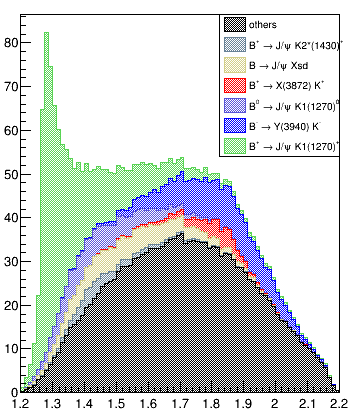}
  %\label{fig:test2}
\end{minipage}%
\begin{minipage}{.33\textwidth}
  \centering
  \includegraphics[width=1\textwidth]{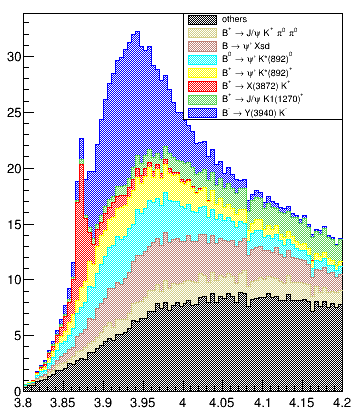}
  %\label{fig:test2}
\end{minipage}%
\caption{ Backgrounds in $M_{J/\Psi\omega}$ and $M_{\omega K}$ for neutral $B$ meson are plotted in left and middle plot  respectively. Right side one is  representing the $M_{J/\Psi\omega}$ for whole MC sample after applying $\Psi^\prime K^*$ veto and $M_{\omega K}$ cut. }
	\end{figure}
	We use large $B\rightarrow J/\Psi X$ inclusive MC sample (having 100 times  statistics compared to data) to understand the sources of background. As we are interested in $M_{J/\Psi\omega}$, we check the distribution for $M_{J/\Psi\omega}$  and $M_{\omega K}$ (Fig1). One can clearly see from the $M_{\omega K}$ distribution that, by applying a cut $M_{\omega K}>1.4$ GeV most of background coming from  $B\rightarrow J/\Psi K_1(1270)$ decay can be removed. For extracting $X(3872)$ and $X(3915)$ signal we look at $M_{J/\Psi\omega}$ distribution from 3.81 to 4.2 GeV$/c^2$.
	
\subsection{$\Psi^\prime K^*$ veto}
  Background coming from the $B\rightarrow \Psi^\prime K^*$ decay is peaking around the signal peak. Here we expect $J/\Psi\pi^+\pi^-$ coming from $\Psi^\prime$ and $\pi^0K$ from $K^*$ to mimic our signal. Therefore, we apply $\Psi^\prime K^*$  veto, $3.67\rm{GeV/c}^2 <M_{J/\Psi\pi^+\pi^-}< 3.72\rm{GeV/c}^2$ and $0.79 \rm{GeV/c}^2 <M_{\pi^0K}< 0.99 \rm{GeV/c}^2$ to reduce such background.
  
\section{Signal extraction with maximum likelihood fit}
\begin{figure}
\centering
\begin{minipage}{.5\textwidth}
  \centering
  \includegraphics[width=1\textwidth]{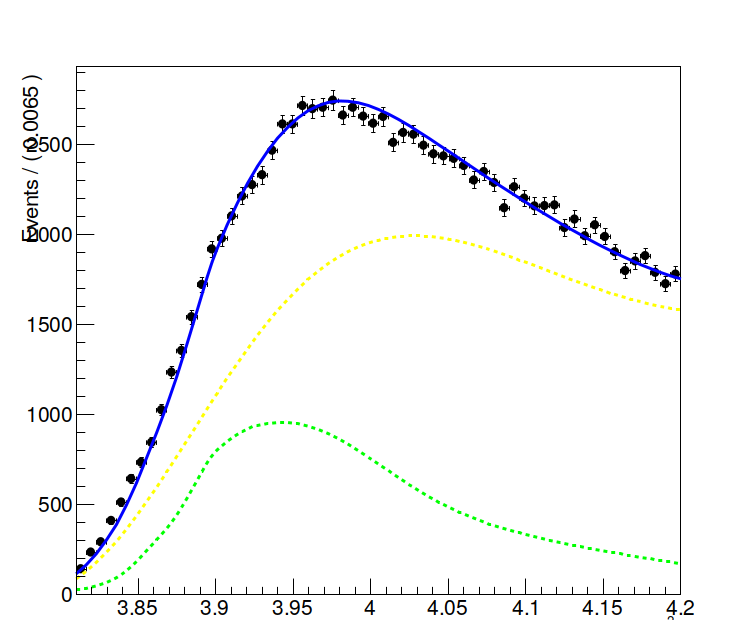}
   %\caption{$K^\pm$ events}
  %\label{fig:test1}
\end{minipage}%
\begin{minipage}{.5\textwidth}
  \centering
  \includegraphics[width=1\textwidth]{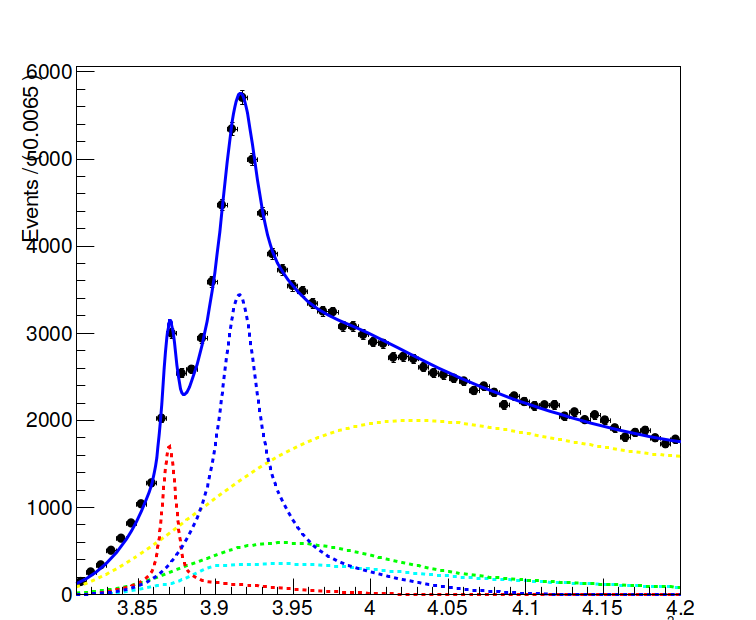}
  %\label{fig:test2}
\end{minipage}%
\caption{1D UML fit to $M_{J/\Psi\omega}$ distribution for the possible background estimated from $B\rightarrow J/\Psi X$ sample (left) and total fit after signal  inclusion for $B^\pm\rightarrow J/\Psi\omega K^\pm$ decay modes (right).}
\end{figure}
For extracting signal efficiency we perform 1D unbinned maximum likelihood fit (UML) for $X(3872)$ and $X(3915)$ with signal MC sample. We model each of the signals with one Gaussian and two bifurcated Gaussians. For the peaking backgrounds, $B^\pm \rightarrow \Psi^\prime {K^*}^\pm$ and $B^0 \rightarrow \Psi^\prime {K^*}^0$, we use one Gaussian and two bifurcated Gaussian. Rest of the backgrounds have flat nature in signal region. Therefore, we use threshold function to model those backgrounds. Finally, we combine all the PDFs in a single PDF fixing all the parameters from signal MC including mean and sigma for $X(3872)$ and $X(3915)$, floating the yields of all three PDFs (Fig. 2).   

\section{Conclusion}
A preliminary MC study for $B\rightarrow J/\Psi\omega K$ is presented here. We tried different methods to reduce the cross feed and to improve the resolution of $M_{J/\Psi\omega}$. Precise measurement of $\mathcal{B}[X(3872)\rightarrow J/\Psi\omega]$ to $\mathcal{B}[X(3872)\rightarrow J/\Psi\pi^+\pi^-]$ will help in understanding the nature of $X(3872)$.


\begin{thebibliography}{99}
\bibitem{conjugation}
Charge-conjugate modes are included throughout the analysis unless stated otherwise.
\end{thebibliography}
\end{document}